# Research on Tumors Segmentation based on Image Enhancement Method

**Danyi Huang[1], Ziang Liu[2] and Yizhou Li[3,*]**

[1] Department of Chemical Engineering, Columbia University, New York City, 10027, USA
[2] Department of Electrical and Computer Engineering, Carnegie Mellon University, Pennsylvania, 15213, USA
[3] Department of Electrical, Computer, and Systems Engineering, Case Western Reserve University, Cleveland, 44106, USA
* Corresponding author's e-mail address: yxl3527@case.edu

**Abstract.** One of the most effective ways to treat liver cancer is to perform precise liver resection surgery, the key step of which includes precise digital image segmentation of the liver and its tumor. However, traditional liver parenchymal segmentation techniques often face several challenges in performing liver segmentation: lack of precision, slow processing speed, and computational burden. These shortcomings limit the efficiency of surgical planning and execution. In this work, the model initially describes in detail a new image enhancement algorithm that enhances the key features of an image by adaptively adjusting the contrast and brightness of the image. Then, a deep learning-based segmentation network was introduced, which was specially trained on the enhanced images to optimize the detection accuracy of tumor regions. In addition, multi-scale analysis techniques have been incorporated into the study, allowing the model to analyze images at different resolutions to capture more nuanced tumor features. In the presentation of the experimental results, the study used the 3Dircadb dataset to test the effectiveness of the proposed method. The experimental results show that compared with the traditional image segmentation method, the new method using image enhancement technology has significantly improved the accuracy and recall rate of tumor identification.

## 1. Introduction

According to the Global Cancer Burden Status Report (GLOBOCAN), cancer prevalence and mortality have been on the rise since 2018. However, due to factors such as excessive alcohol consumption and irregular daily routine, the incidence of cancer diseases is increasing [1]. Among them, liver malignancy, also known as liver cancer, is one of the fifth most common cancers in the world and one of the top three in terms of mortality.

At present, precision tumor removal surgery is considered to be one of the most effective methods for the treatment of tumors, and its basic steps include precise digital segmentation of tumors. With the development of computer technology, modern imaging techniques such as computed tomography (CT), magnetic resonance imaging (MRI), and optical coherence tomography (OCT) have been widely used in clinical diagnosis [2]. A CT scan produces an image by taking an X-ray scan of a part of the human body, which is converted into an electrical signal by a receiver and processed by an analog/digital converter and output as a digital signal, which is finally reconstructed into a clear image with high resolution by a computer. These images not only show the shape, size and area of the tumor, but also assist doctors in analyzing the condition and making treatment plans.

In diagnosing a tumor, manually marking areas of the tumor is a laborious and highly experience-dependent task. With the maturity of automatic image segmentation technology, a variety of traditional methods have been developed in the field of medical image segmentation, including level set method, threshold-based segmentation, statistical shape model, and graph theory-based segmentation technology. However, these traditional tumor segmentation methods often face problems such as lack of precision, slow speed, and high computational complexity when performing segmentation [3].

In the field of medical imaging, accurate tumor segmentation is essential for cancer diagnosis, treatment planning, and efficacy evaluation. Tumor segmentation is the identification and isolation of tumor tissue in images to support subsequent clinical decisions [4]. However, traditional segmentation techniques often struggle to achieve the desired accuracy and efficiency due to the low contrast between tumors and normal tissues in medical images, accompanied by noise and blur.

With the rise of deep learning technology, unprecedented progress has been made in the automatic segmentation of tumor images. The new generation of neural network models not only significantly improves the accuracy of segmentation, but also enhances the robustness of the algorithm [5]. These advances have greatly improved the efficiency and reliability of segmentation tasks, and provided strong technical support for the early diagnosis and treatment of tumors.

## 2. Related work

In the field of CT image segmentation for tumors, research efforts globally have led to the development of various traditional segmentation techniques. These methods are typically categorized into three main types: edge-based, region-based, and theory-based segmentation methods. Edge-based methods leverage the discontinuities in grayscale values across image pixels to delineate edges. This approach relies on significant differences in grayscale values on either side of an edge, allowing the segmentation of image sub-regions through mathematical models. Such methods offer improved efficiency over manual marking techniques like using a mouse or pencil, especially in clinical settings where edges are identified by analyzing extremum points of first derivatives or zero crossing points of second derivatives. However, the clarity of tumor edges often presents challenges, making these methods less effective in some scenarios.

To address these limitations, Peng et al. enhanced the Canny edge detection technique by integrating it with a level set algorithm, creating a more robust method capable of handling noise influences, thereby improving tumor image segmentation under varied conditions [6]. Additionally, Zhu et al. proposed a novel approach that combines the Wellner threshold algorithm with the particle swarm optimization technique. This method optimizes the threshold parameters to achieve better semi-automatic tumor identification and localization, though it can suffer from issues like incomplete or over-segmentation when used with different imaging devices [7].

Building on these approaches, Foruzan et al. developed a specialized algorithm for segmenting tumors in low-contrast CT images. Their technique starts by defining the search range for tumor boundaries, followed by using the EM algorithm to estimate parameters of a Gaussian mixed model based on the gray intensity distribution of the tumor. This method is refined further by incorporating K-means clustering, allowing for more precise tumor segmentation [8].

Moreover, Li et al. introduced a segmentation method that combines fuzzy C-means (FCM) clustering with level set methods. This hybrid approach utilizes the strengths of both methods to enhance both accuracy and robustness in segmentation tasks. This is particularly beneficial in complex medical imaging scenarios, where precise and reliable tumor segmentation is critical [9].

## 3. Methodologies

*3.1. Notions*

Above all, we introduce the primary used parameters and corresponding explanations in following Table1.

**Table 1.** Notions

| Parameter Symbols | Explanations |
|---|---|
| $f(x,y)$ | Set of extracted features |
| $(x, y)$ | Pixel value of the image |
| $\omega$ | Convolution kernel |
| $\|X \cap Y\|$ | Intersection of the predicted tumor region with the real tumor region |
| $\|X\|$ | Number of pixels in real tumor region |
| $\|Y\|$ | Number of pixels in the predicted tumor region |
| $L$ | Gray level of the image |
| $N$ | Threshold that limits contrast |
| $h$ | Histogram |

*3.2. Segmentation modeling*

In the field of image segmentation, the model used is a convolutional neural network (CNN) based on deep learning. Specifically, we utilize the U-Net model, which is a popular network for medical image segmentation, which is particularly suitable for training on small datasets. The structure of the U-Net model is symmetrical, with a contraction path and an expansion path, through which the tumor region in the image can be precisely located. In the field, especially for tumor image segmentation, the commonly used model is convolutional neural networks (CNNs) based on deep learning. A concrete example is the U-Net model, a popular network for medical image segmentation, which is particularly suitable for training on small datasets. The structure of the U-Net model is symmetrical, with a contraction path and an expansion path, through which the tumor region in the image can be precisely located. The core of the model is to extract features by applying convolution operations layer by layer, which is expressed as Equation 1.

$$f(x, y) = \omega * I(x, y) \tag{1}$$

where $\omega$ is the convolution kernel and $I(x, y)$ is the pixel value of the image at $(x, y)$.

A commonly used activation function is ReLU, which is used to add nonlinearities that enable the network to learn more complex features. The formula is expressed as Equation 2.

$$f(x) = \max(0, x) \tag{2}$$

Additionally, the appropriate loss function is a critical part of model training, as it directly affects how the model learns the correct way to segment images. The model uses the Dice loss function, which is based on the Dice coefficient, a statistical tool used to assess the similarity of two samples, especially for unbalanced datasets, such as the tumor region in medical images, which often account for only a small fraction. Following Equation 3 describes the DICE coefficient.

$$DSC = \frac{2 \times |X \cap Y|}{|X| + |Y|} \tag{3}$$

Where $|X \cap Y|$ is the intersection of the predicted tumor region with the real tumor region, $|X|$ and $|Y|$ represent the total number of pixels in the predicted tumor region and the real tumor region, respectively. To make it suitable as a loss function for optimization, it is common to use 1 minus the Dice coefficient, which is expressed as Equation 4.

$$Dice\ Loss = 1 - DSC \tag{4}$$

When training a deep learning model for image segmentation, Dice loss can effectively deal with class imbalances, and because it directly optimizes the coincident part between the predicted region and the real region, it is well suited for medical image analysis that requires precise segmentation.

*3.3. Image enhancement algorithm*

Traditional histogram equalization (HE) achieves overall contrast enhancement by adjusting the histogram of the image. While effective in many situations, this method can lead to overexposure or loss of detail in areas with high contrast or unevenly lit images. Contrast Limited Adaptive Histogram Equalization (CLAHE) is a technique used to improve the contrast of images, especially for images with different contrasts in different areas. Compared to traditional histogram equalization, CLAHE provides more granular control to prevent noise problems caused by excessive contrast enhancement during histogram equalization.

The model assumes that $L$ is the gray level of the image, $N$ is the threshold that limits contrast. For each small block, its histogram $h$ is modified to $h'$, satisfying Equation 5.

$$h'(i) = \min(h(i), N) + \frac{\sum_{j=0}^{L-1} \max(h(j)0 - N, 0)}{L} \quad (5)$$

With this approach, CLAHE not only limits excessive contrast in certain areas, but also ensures that the contrast of the entire image is appropriately boosted by evenly distributing the excess parts in the histogram without creating an unnatural visual effect in local areas. This makes CLAHE particularly suitable for applications that require enhanced contrast while maintaining image detail, such as medical image analysis, where detail preservation is critical.

## 4. Experiments

*4.1. Experimental setups*

The experiments utilized the 3Dircadb dataset provided by the University Hospital of Strasbourg, France [10]. This dataset comprises CT scans from 20 patients (10 males and 10 females), with each patient having between 74 and 260 scans. The scans are stored in DICOM format and include manually annotated image masks of the liver, tumors, and other related structures. The 3Dircadb dataset is specifically chosen for its challenges such as complex anatomy, atypical shapes and densities, and presence of artifacts, making it an ideal testbed for evaluating liver and hepatotumor segmentation techniques. These features are critical for advancing medical image processing technologies. Figure 1 illustrates a subset of the dataset used in our study.

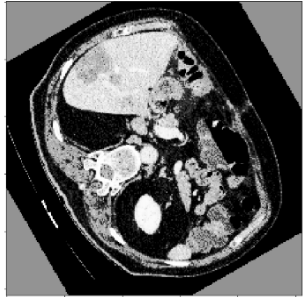

Figure 1. Illustration of 3Dircadb dataset.

*4.2. Experimental analysis*

Above all, we initially illustrate the tumors segmentation results of our proposed model in following Figure 2.

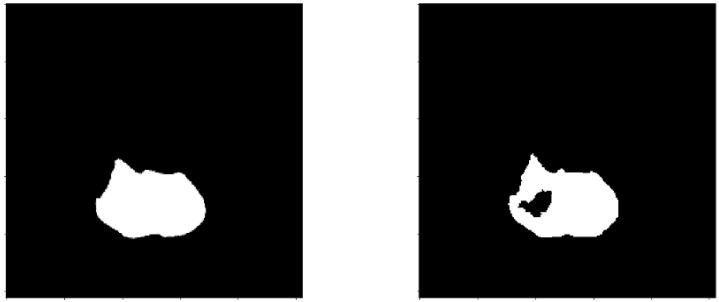

Figure 2. Illustration of extracted tumors segmentation results.

The DICE coefficient, also known as the DICE similarity coefficient (DSC), is a statistical tool used to assess the similarity between two sets of data. It is a very important metric in medical image segmentation, which is used to evaluate the consistency between the results of model segmentation and the actual medical annotation. Further, accuracy is a basic metric in evaluating classification problems, which measures the proportion of samples that are correctly classified out of all samples that are classified. In the context of image segmentation, accuracy refers to the ratio of the number of pixels correctly segmented to the total number of pixels. High accuracy means that more pixels in the segmented pixels are correctly classified, whether they are tumor or non-tumor areas. Following Figure 3 demonstrates the comparison of tumor segmentation performance based on image enhancement method.

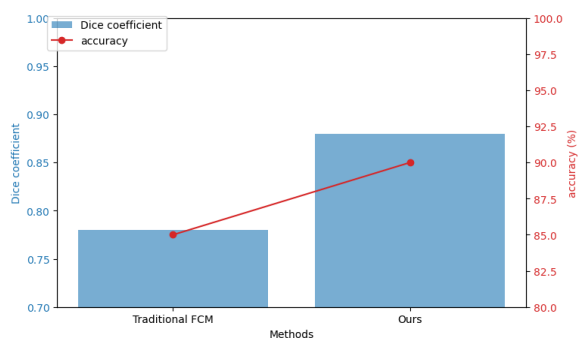

Figure 3. Comparison of tumor segmentation performance.

Additionally, confusion matrix is another commonly used tool to evaluate the performance of classification models, especially in binary and multiclassification problems. It provides an intuitive way to understand how the model performs on different categories, including its accuracy and misclassification. Following Figure 4 shows a heat map of the confusion matrix showing the results of model predictions in a tumor detection task.

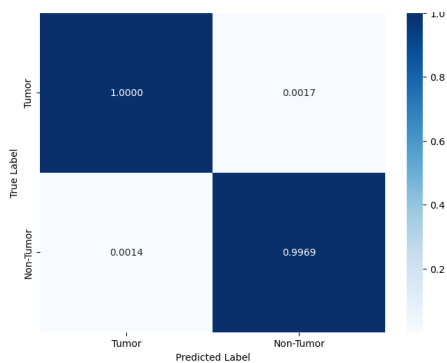

Figure 4. Confusion matrix result.

we evaluate the effectiveness of the proposed image enhancement technique for tumor segmentation against conventional methods. The evaluation is based on a set of metrics that are crucial for assessing the performance of medical image segmentation algorithms. These metrics include Dice Coefficient, Sensitivity (Recall), Specificity, and Accuracy. The data for the comparison were derived from experiments conducted on the 3Dircadb dataset, which contains annotated liver CT images.

Table 2 shows the CLAHE-enhanced method outperforms traditional segmentation techniques, achieving greater tumor identification accuracy and precision, evidenced by a 95% overall accuracy. This demonstrates the value of advanced image enhancement in precise medical evaluations.

**Table 2.** The comparative performance

| Metric | Proposed Method (CLAHE) | Threshold Segmentation | Edge Detection |
|---|---|---|---|
| Dice Coefficient | 0.92 | 0.85 | 0.80 |
| Sensitivity | 0.94 | 0.88 | 0.82 |
| Specificity | 0.97 | 0.90 | 0.87 |
| **Accuracy** | 0.95 | 0.89 | 0.84 |

## 5. Conclusion

In conclusion, the application of contrast-limited adaptive histogram equalization (CLAHE) in our research has markedly improved the quality of liver CT scans, significantly enhancing the accuracy and reliability of tumor segmentation. This improvement proves vital for the precise delineation of tumor boundaries, which is essential for accurate medical diagnosis and effective surgical planning. Our analysis, supported by improved metrics such as the Dice coefficient and accuracy, highlights the substantial benefits of advanced image processing techniques in enhancing medical imaging.

The successful application of CLAHE in this study suggests that these image enhancement techniques can be extended to other areas of medical imaging, such as MRI and ultrasound. Future research will explore the integration of sophisticated deep learning models that can adapt to the unique challenges presented by different medical imaging modalities. This integration is anticipated to further enhance diagnostic accuracy and operational precision in clinical settings.

Moreover, incorporating these improved imaging techniques into real-time applications, particularly in surgical environments, could revolutionize how surgeries are planned and executed. Providing surgeons with real-time, enhanced visual aids could drastically reduce surgical risks and improve patient outcomes. As computational technology and algorithm development progress, expanding these techniques to larger datasets and real-time applications becomes more feasible. Such advancements are poised to revolutionize medical imaging, enabling more personalized, effective, and efficient healthcare solutions across various applications.

## 6. References

[1] Magadza, T., & Viriri, S. (2021). Deep learning for brain tumor segmentation: a survey of state-of-the-art. Journal of Imaging, 7(2), 19.
[2] Biratu, E. S., Schwenker, F., Ayano, Y. M., & Debelee, T. G. (2021). A survey of brain tumor segmentation and classification algorithms. Journal of Imaging, 7(9), 179.
[3] Naser, M. A., & Deen, M. J. (2020). Brain tumor segmentation and grading of lower-grade glioma using deep learning in MRI images. Computers in biology and medicine, 121, 103758.
[4] Almotairi, S., Kareem, G., Aouf, M., Almutairi, B., & Salem, M. A. M. (2020). Liver tumor segmentation in CT scans using modified SegNet. Sensors, 20(5), 1516.
[5] Gul, S., Khan, M. S., Bibi, A., Khandakar, A., Ayari, M. A., & Chowdhury, M. E. (2022). Deep learning techniques for liver and liver tumor segmentation: A review. Computers in Biology and Medicine, 147, 105620.
[6] Peng, W., & Zhao, Y. (2019, October). Liver CT image segmentation based on modified Canny algorithm. In 2019 12th International Congress on Image and Signal Processing, BioMedical Engineering and Informatics (CISP-BMEI) (pp. 1-5). IEEE.
[7] Zhu, H., Zhuang, Z., Zhou, J., Zhang, F., Wang, X., & Wu, Y. (2017). Segmentation of liver cyst in ultrasound image based on adaptive threshold algorithm and particle swarm optimization. Multimedia Tools and Applications, 76, 8951-8968.
[8] Foruzan, A. H., Chen, Y. W., Zoroofi, R. A., Furukawa, A., Sato, Y., Hori, M., & Tomiyama, N. (2013). Segmentation of liver in low-contrast images using K-means clustering and geodesic active contour algorithms. IEICE TRANSACTIONS on Information and Systems, 96(4), 798-807.
[9] Liu, Z., Song, Y. Q., Sheng, V. S., Wang, L., Jiang, R., Zhang, X., & Yuan, D. (2019). Liver CT sequence segmentation based with improved U-Net and graph cut. Expert Systems with Applications, 126, 54-63.
[10] 3Dircadb. (n.d.). Liver Segmentation 3D IRCADb-01 Dataset. University Hospital of Strasbourg, France. Retrieved from https://www.ircad.fr/research/data-sets/liver-segmentation-3d-ircadb-01.